\documentclass{aa}     
\usepackage{epsfig}
\def\a85{ABCG~85{}}
\def\a87{ABCG~87{}}
\def\a89{ABCG~89{}}

\def\dip{``Dip''{}}
\def\fg{``Foreground Group''{}}
\def\kms{km~s$^{-1}${}}
\def\deg{$^\circ $}
\thesaurus{03 (11.03.1; 11.03.4; 12.04.1; 13.25.2) }

\title{The rich cluster of galaxies ABCG 85. III. Analyzing the 
ABCG 85/87/89 complex. 
\thanks{Based on ROSAT Archive data and on observations collected at the 
European Southern Observatory, La Silla, Chile}}

\author {F.~Durret\inst{1,2}\and W.~Forman\inst{1,3} \and 
D.~Gerbal\inst{1,2}  \and C.~Jones\inst{3} \and A.~Vikhlinin\inst{3}}

\offprints{ }
\institute{
Institut d'Astrophysique de Paris, CNRS, Universit\'e Pierre et
Marie Curie, 98bis Bd Arago, F-75014 Paris, France
\and
        DAEC, Observatoire de Paris, Universit\'e Paris VII, CNRS (UA 173),
F-92195 Meudon Cedex, France
\and
Harvard-Smithsonian Center for Astrophysics, 60 Garden Street, Cambridge,
MA 02138, USA
}
\date{Received, 1997; accepted,}
\begin{document}

\maketitle

\begin{abstract}
We present a combined X-ray and optical analysis of the ABCG 85/87/89
complex of clusters of galaxies, based on the ROSAT PSPC image,
optical photometric catalogues (Slezak et al. 1998), and an optical
redshift catalogue (Durret et al. 1998).  From this combined data
set, we find striking alignments at all scales at
PA$\simeq$160\deg. At small scales, the cD galaxy in ABCG 85 and the
brightest galaxies in the cluster are aligned along this PA. At a
larger scale, X-ray emission defines a comparable PA south-southeast
of ABCG 85 towards ABCG 87, with a patchy X-ray structure very
different from the regular shape of the optical galaxy distribution in
ABCG 87. The galaxy velocities in the ABCG 87 region show the
existence of subgroups, which all have an X-ray counterpart, and
seem to be falling onto ABCG 85 along a filament almost perpendicular
to the plane of the sky.

To the west of ABCG 85, ABCG 89 appears as a significant galaxy
density enhancement, but is barely detected at X-ray wavelengths. The
galaxy velocities reveal that in fact this is not a cluster but two
groups with very different velocities superimposed along the line of
sight. These two groups appear to be located in intersecting sheets on
opposite sides of a large bubble.

These data and their interpretation reinforce the cosmological
scenario in which matter, including galaxies, groups and gas, falls
onto the cluster along a filament.

\keywords{Galaxies: clusters: general; Clusters: individual: ABCG 85; 
X-rays: galaxies}
\end{abstract}

\section{Introduction}

One of the major changes in our understanding of our universe has been
the realization of the rich and complex structures which are apparent
in the large scale distribution of galaxies. From both redshift
surveys and projected galaxy distributions, the appearance of galaxy
voids and supercluster filaments has become clear. The large scale
structure also is an important constraint for different cosmological
scenarios. On smaller scales, investigations of the relationships
between nearby galaxy clusters suggest that clusters retain
information about the large scale structures from which they form
(e.g., van Haarlem \& van de Weygaert 1993; West et al. 1995, West
1997; Colberg et al. 1997). The frequency of substructure may also
provide constraints on $\Omega$ (Richstone et al. 1992, Mohr et
al. 1995, Buote \& Xu 1997; see also Kauffmann \& White 1993).

As part of a survey for substructure, we have studied the large scale
mass distribution around the richness class~1 cluster ABCG 85 (Abell
et al. 1989), using X-ray and optical observations. As a bright,
luminous, relatively nearby cluster, ABCG 85 has been studied
extensively. At a redshift of $z=0.0555$, the angular scale for ABCG
85 corresponds to $72~h_{50}^{-1}$~kpc arcmin$^{-1}$ (we assume H$_0$=
50 km sec$^{-1}$ Mpc$^{-1}$ throughout). ABCG 85 contains a cD galaxy
close to its center.  The cluster X-ray characteristics include a
peaked emission profile, harboring a cooling flow, emission from
individual cluster member galaxies, an X-ray emitting subcluster south
of the cluster center (hereafter the south blob), a superposed
foreground group of galaxies (to the west-northwest), as well as
additional foreground and background structures detected from the
optical spectroscopic observations (see Jones et al. 1979, Pislar et
al.  1997, Lima-Neto et al. 1997, Slezak et al. 1998, Durret et
al. 1998, and in preparation for detailed discussions of the X-ray and optical
observations).

In this contribution, we report on the X-ray and optical properties of
a larger region around ABCG 85 than has been considered in previous
studies, and in particular include the nearby clusters ABCG 87 and
ABCG 89. We find that ABCG 85 itself exhibits preferential alignments
on scales from 100 kpc to $\sim 4$Mpc (in projection on the sky). In
particular, the cluster cD galaxy is elongated along the same position
angle as a large filamentary structure in X-rays roughly coinciding
with the optical position of ABCG 87. Located to the east of ABCG 85
is ABCG 89, not detected as an extended X-ray source. We also address
why ABCG 87 appears different in the optical and X-ray and why ABCG 89
appears as a cluster in the optical observations and not in X-rays.

Based on these results, we discuss the ABCG 85/87/89 complex and
describe the alignments and/or structures in the context of a large
scale structure formation scenario such as that proposed by West et
al. (1995; see also van Haarlem \& van de Weygaert 1993 and Colberg
et al. 1997).  In Sect. 2 we present the X-ray imaging analysis and
galaxy distributions. Sect. 3 discusses in detail the galaxy
velocity distribution. A model is proposed in Sect. 4 and discussed
in Sect. 5.

\begin{figure}[tbp]
\centerline{
        \psfig{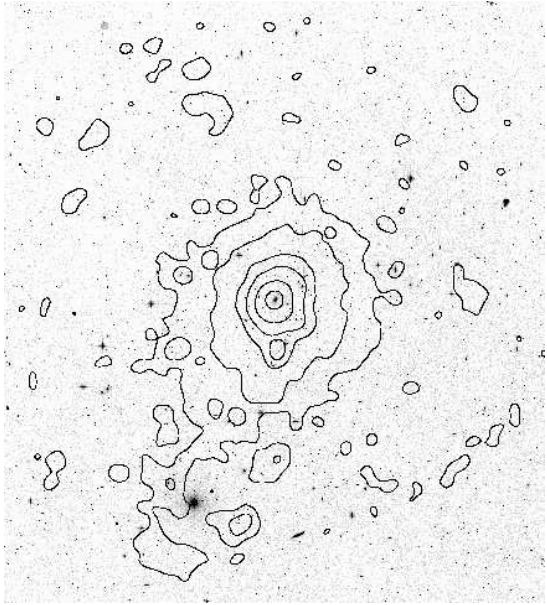}}
\caption{Optical digitized sky survey of the ABCG 85 region, with
the isophotes of the ROSAT PSPC image of a flat-fielded $55'$ radius 
field  superimposed. The energy band used is 0.4-2.0 keV and the image has 
been smoothed with a $60''$ Gaussian after flat-fielding. The cD galaxy at the
X-ray cluster center (peak of the X-ray emission) is clearly
visible on the optical map; a bright star is located in the region of extended
X-ray emission, but it does not significantly contribute to
the observed X-ray emission.}
\protect\label{xcont}
\end{figure}

\section{X-ray Imaging Analysis and Galaxy Distribution}\label{data}

\subsection{ROSAT PSPC X-ray Observations of ABCG 85}

We have analyzed the ROSAT PSPC observations of ABCG 85 using an
image in the energy band 0.5-2.0 keV binned into 15 arcsecond pixels
and flat fielded. Figure \ref{xcont} shows the resulting ROSAT PSPC
image of the $55'$ radius field centered on ABCG 85, smoothed with a
$60''$ Gaussian to enhance the signal to noise ratio at large
off-angle distances. The image clearly shows emission peaked on the cD
at the cluster center, a second region of extended emission (the
south blob) approximately $10'$ south of ABCG 85 (720 kpc, projected
on the sky), as well as point sources scattered over the PSPC field of
view (FOV).  In addition to these features, there is an excess of
emission towards the south south-east, extending almost to the edge of
the FOV.  The emission to the southeast is concentrated in a narrow
range of position angles, approximately between 135$^\circ$ and
180$^\circ$ (counter clockwise from north); the positions and net
counts of individual sources from this region are listed in Table~1.
We also generated an image following the Snowden et
al. (1994; see also Snowden 1995) prescription which slightly reduces
the field of view but shows the same features in the overlapping regions.

\begin{table}
\begin{center}

\caption{X-ray Source Properties}

\begin{tabular}{rrrr}
\hline
ID & $\alpha$~~~~~ & $\delta$~~~~ & Net~  \\
   & \multicolumn{2}{c}{(J2000.0)} & counts  \\
\hline
1 & 0h 42mn 25s & -9\deg 48'  & 47.7    \\
2 & 0h 42mn 56s & -10\deg 00' & 48.1   \\
3 & 0h 43mn 51s & -9\deg 51'  & 46.0    \\
4 & 0h 43mn 51s & -10\deg 06' & 47.4  \\
\hline
\end{tabular}
\end{center}
\end{table}

As shown in Table~1, the sources are all of comparable intensity. If we
assume a typical source contains 45 net counts, we find a flux of
$4.8\times10^{-14}$ ergs cm$^{-2}$ s$^{-1}$ (using a Galactic column
density of $3.0\times10^{20}$ cm$^{-2}$ and a spectrum characterized
by a temperature of 1.4 keV) for an exposure time of 9.8 ksec. At
the distance of ABCG 85, this corresponds to a luminosity of
$9\times10^{41}$ ergs s$^{-1}$. These luminosities are typical of
those of small groups of galaxies (e.g., Henry et al. 1995, Pildis et
al. 1995, Ponman et al. 1996).

The azimuthal distribution computed by binning the counts in the PSPC
image in 45\deg\ sectors over the radial range from 27.5 to 55.0
arcminutes (110-220 $15''$ pixels) centered on the peak of the surface
brightness distribution, shows a clear excess of approximately 500 cts
(2750 compared to 2250 cts) compared to the mean at other azimuths.
The smaller scale sources contain only approximately 50\% of this
total excess, indicating additional flux on larger scales (clearly
seen in Figs. 2 and 3, see below).
This excess emission (excluding the four concentrations in Table~1) has
a flux corresponding to $2.7\times10^{-13}$ ergs cm$^{-2}$ s$^{-1}$,
assuming it has the same spectral properties as we assumed for the
individual sources above. While our assumption for the sources may be
reasonable, since their luminosities are equivalent to those of
groups, the excess flux could be characterized by a significantly
cooler temperature if it lies in still weaker potentials than typical
of small groups.

\begin{figure}[tbp]
        \centerline{
\psfig{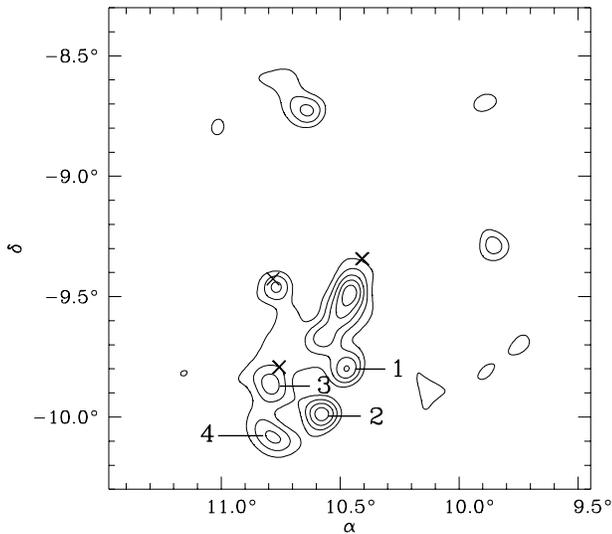}
}
        \caption{Residual X-ray emission obtained after subtracting the
smooth emission derived from the north half of the image, as described in
the text. The crosses indicate the centers of the three clusters ABCG
85, 87 and 89. The numbers correspond to the sources defined in Table 1.}
\protect\label{excessx}
\end{figure}

\begin{figure}[tbp]
        \centerline{
\psfig{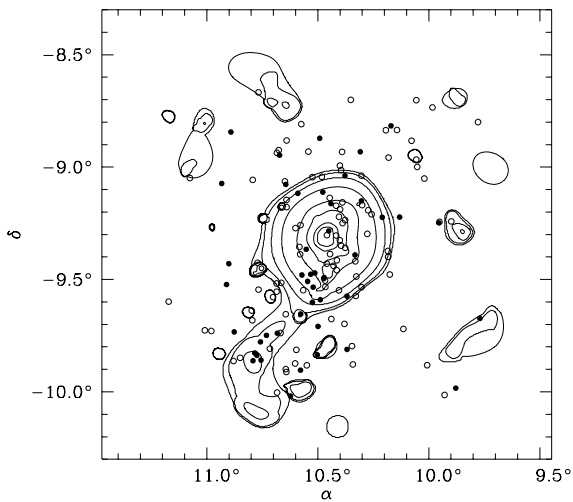}
}
        \caption{X-ray contour map after wavelet analysis, onto which
are superimposed the galaxies with magnitudes brighter than ${\rm b_J}=17$, and
belonging to the ABCG 85 cluster. The coding is as follows: black circles: 
galaxies in the [14500-16100 \kms] velocity range, white circles: 
galaxies in the [16100-18500 \kms] velocity range.}
\protect\label{wavelets}
\end{figure}

We generated an image of the excess X-ray emission to the south-east
of ABCG 85. We built a synthetic image for the whole cluster, using
the azimuthal profile of the northern half of ABCG 85, which was then
subtracted from the original data (and normalized by dividing by the
symmetric model). The result is displayed as a contour map in
Fig.~\ref{excessx}. 
The  surface
brightness enhancement which initially extends to the south from the
center of A85, shifts in angle towards the southeast.

A wavelet transform analysis, which allows
extraction of emission at different spatial scales, gives comparable
results (Fig.~\ref{wavelets}).

\subsection{The distribution of galaxies in the ABCG 85 Region}

Slezak et al. (1998) compiled two photometric catalogues, one of
11862 galaxies in a $5\times 5^\circ$ field, and one of 4232 galaxies
in a 1 degree radius region centered on the cluster to a limiting
magnitude of b$_{\rm J} < 21.3 $; the latter is 95\% complete to 
b$_{\rm J}< 19.75$. 

By using the first and second moments of the distribution of galaxy
positions, we calculated the ellipticity and the major axis position
angle of the galaxy distribution; the ellipticity is e=0.82 and the
major axis is along PA$\simeq$160\deg.  This alignment is also
illustrated by the brightest cluster galaxies (11 galaxies brighter
than R=15 and having redshifts belonging to the cluster) which are
distributed in a narrow band with PA$\simeq$135\deg. We also note that
the cD galaxy itself is elongated with PA$\simeq$152\deg.

\begin{figure}[tbp]
        \centerline{
        \psfig{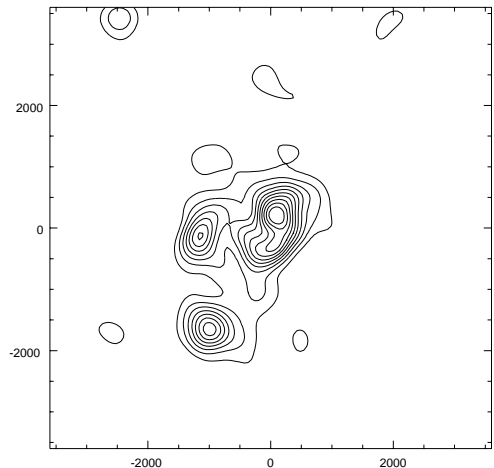}}
        \caption{Adaptive kernel map of the galaxy distribution, drawn
	for galaxies brighter than 
        B$_{\rm J}=19.75$ in a field of $2\times 2$ square degrees
        (2018 objects). The initial kernel chosen was half the Silverman
        value and 10 bootstrap resamplings were performed.}
\protect\label{kernel}
\end{figure}

We used the optical photometric catalogues of Slezak et al. to
investigate the structure corresponding to the X-ray emission.  A
kernel map (Silverman 1986) showing the smoothed galaxy number density
is displayed in Fig.~\ref{kernel}. This figure shows
the alignment at PA$\simeq$160\deg\ of the contours extending to the
south east from the center of ABCG 85.  The prominent azimuthally
symmetric peak located $\sim33'$ southeast of the cD galaxy at the
position of ABCG 87 also defines the same $\simeq$160\deg\ PA. Another
structure is seen at the optical position of ABCG 89. A comparison of
the X-ray and optical images (Figs.~\ref{excessx}, \ref{wavelets} and
\ref{kernel}) is instructive.  All the features found in the optical
at PA$\simeq$160\deg\ are  seen in X-rays, with emission extending
from ABCG 85 southeast toward ABCG 87.

Thus both the X-ray emission and the optical galaxies show an excess
of matter in a large region southeast of ABCG 85. Nearly half of the
X-ray emission arises from a few sources, reminiscent of groups of
galaxies.

We find that galaxies with velocities smaller than the cluster
bi-weight (BWT; Beers et al. 1990) mean ($v_{\rm
BWT}=17546$ \kms) and smaller than the velocity of the central cD
($v_{\rm cD}=16734$ \kms) are concentrated in two regions along the
southern filamentary extension (see Fig.~\ref{wavelets}).  These
galaxies have velocities in the 14500-16100 \kms\ range and also lie
at the same position angle as the previously mentioned structures.

A major difference in the X-ray and optical maps is in the appearance
of the region directly to the east of ABCG 85. No extended X-ray
emission is found corresponding to the prominent galaxy concentration
at the location of ABCG 89. Also, the X-ray appearance at the position
of ABCG 87 is not consistent with the presence of a regular, relaxed
system, as is suggested by the optical.  The X-ray structure is
elongated and shows more substructure/clumpiness. On the contrary, a
relaxed system would show X-ray emission having less ellipticity than
found in the optical (see e.g. Fabricant et al. 1984).

We now investigate the three dimensional galaxy distribution (space
and velocity) to clarify the differences and similarities seen in the
X-ray and optical.

\section{Galaxy velocity distribution}\label{carottage}

We use the extensive spectroscopic catalogues of galaxies in this
region to analyse several fields around each cluster with the goal of
determining the three dimensional  structure of the ABCG 85/87/89
complex.  The catalogue by Durret et al. (1998) includes 551 galaxy
velocities in the direction of ABCG 85; among these, 305 galaxies lie
in the velocity range 13000--20000 \kms, which are assumed to belong
to ABCG 85.  We also used five galaxy velocities for ABCG 87 from the
ENACS catalogue (Katgert et al. 1997).

To analyse the distribution of velocities, we use a wavelet
reconstruction as described in Fadda et al. (1997). This method
provides velocity density profiles. The calculation of significance
for features derived in the wavelet reconstruction is described in
their paper. Mathematically, it is possible to perform analyses to very
small velocity scales; however, the features found may not be significant. 
We will therefore use only the largest scales.  
Notice that even if we are not able to be very confident in each peak
for each sample, the fact that the same velocity feature is observed
in adjacent zones, using independent velocity samples, argues in
favour of its true existence.

\begin{table}[h!]
        \centering
        \caption{Characteristics of the seven samples.}
        \begin{tabular}{l c c r r}
\hline
                Name & $\alpha$ & $\delta$ & Radius & Number  \\
   & \multicolumn{2}{c}{(J2000.0)} & ('') & of \\
   &  &  &     & galaxies \\
\hline
        A85 (Sc)  & 0h 41mn 51.9s & -9\deg 18' 17'' & 850 & 123 \\
        A89     & 0h 43mn 08.0s & -9\deg 26' 35'' & 550 &  43 \\
        A87     & 0h 43mn 00.0s & -9\deg 49' 00'' & 850 &  60 \\
        C89     & 0h 40mn 38.0s & -9\deg 09' 59'' & 850 &  43 \\
        C87     & 0h 40mn 40.0s & -8\deg 47' 34'' & 850 &  40 \\
        C1      & 0h 40mn 44.3s & -9\deg 41' 37'' &1000 &  40 \\
        C2      & 0h 42mn 59.5s & -8\deg 54' 47'' & 900 &  44 \\
\hline
        \end{tabular}
        \label{carottes}
\end{table}

\subsection{Velocity subsamples} 

\begin{figure}[tbp]
        \centerline{
        \psfig{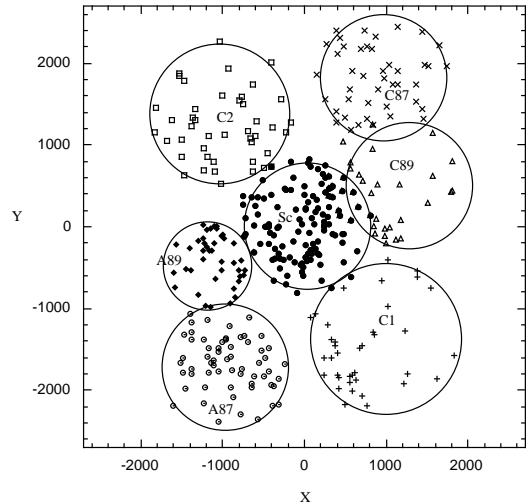}}
        \caption{Distribution of the galaxies into the seven samples.}
\protect\label{echantillon}
\end{figure}

We defined seven subsamples of galaxies chosen in circular regions
on the sky (see Table~\ref{carottes} and Fig.~\ref{echantillon}). One
is located at the center of the cluster ABCG 85 (named Sc) and two
(named A89 and A87) are located at the positions of ABCG 89 and ABCG
87 as indicated by the NED database. Two control subsamples are chosen
at the symmetrical positions of A 87 and A 89 relative to the ABCG 85
center: C87 is symmetric to A87, and C89 to A89. To these five
samples, we add two comparison samples: C1 between A87 and C89 and C2
symmetric to C1 relative to the ABCG 85 center (see
Fig.~\ref{echantillon}). The radii of all these circles have been
chosen in each case to avoid superposition of possible structures,
while obtaining a significant number of velocities in each sample.

\begin{figure}[tbp]
        \centerline{
        \psfig{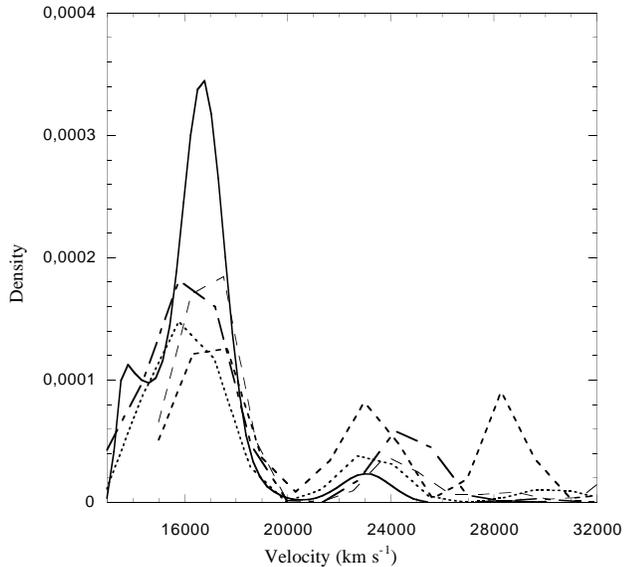}}
    \caption{Velocity distribution for five velocity samples, with the 
following line coding: full line: Sc, short dashed line: A89, dotted line:
A87, long dashed line: C87, dot-dashed line: C89.}
        \protect\label{grandechelle}
\end{figure}

\begin{figure}[tbp]
\vskip 1.0truecm
        \centerline{
        \psfig{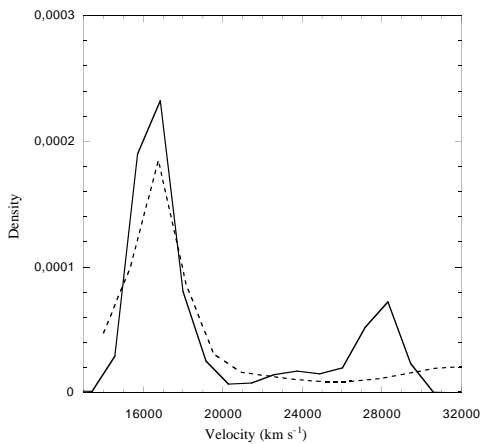}}
    \caption{Velocity distribution for two control velocity samples. 
    Continuous line: C1, dotted line: C2.}
        \protect\label{temoin}
\end{figure}

Velocity clustering is observed between 13000 \kms\ and 32000 \kms, as
illustrated in Fig.~\ref{grandechelle}. The velocity distributions for
the control samples C1 and C2 are displayed in Fig.~\ref{temoin}. The
most prominent feature centered at about 17000 \kms\ corresponds to
the ABCG 85 cluster. In Fig.~\ref{grandechelle}, a second feature is
seen at about 23000 \kms.

\subsection{The 20000 - 32000 \kms\ velocity range} \label{large}

\subsubsection{A galaxy sheet between 20000 \kms\ and 26000 \kms}

The peak observed between 20000 \kms\ and $\sim$26000 \kms\
(Fig.~\ref{grandechelle}) is particularly strong in the region of A89
(Figs.~\ref{grandechelle} and \ref{vitS1}); this peak (hereafter ABCG
89b) has a velocity dispersion $\sigma = 780 \pm 260$ \kms. Such a
value is too large for a single relaxed group, thus
suggesting either substructure or contamination by other structures.
As Fig.~\ref{grandechelle} shows, the maximum between 20000 and 26000
\kms\ is not at the same velocity for all the samples. In particular, the peaks
for the control samples C87 and C89 are higher: $\simeq$24000 \kms\
for C87 and $\simeq$25000 \kms\ for C89.

The clustering of velocities, together with the variation of the peak
velocity, suggest the presence of a sheet of galaxies inclined with
respect to the plane of sky. This sheet is not  visible in the
two control samples C1 and C2. 

\subsubsection{A structure at $\simeq28000$ \kms: ABCG 89c}

A second peak is observed in the A89 velocity distribution at about
28500 \kms\ (see Figs.~\ref{grandechelle} and \ref{vitS1}).  An
enhancement is also found at this velocity in the C1 control sample
(Fig.~\ref{temoin}), but no peak is present in the five other velocity
samples in Fig.~\ref{grandechelle}.  This high redshift component of
A89 has a velocity dispersion of $\sigma$=$333\pm118$ \kms, typical of
a galaxy group or small cluster. Obviously  ABCG 89 is
not a simple cluster, but rather a complicated superposition of
structures. 

\subsection{The 13000 - 20000 \kms\ velocity range} \label{short}

\begin{figure}[tbp]
        \centerline{
        \psfig{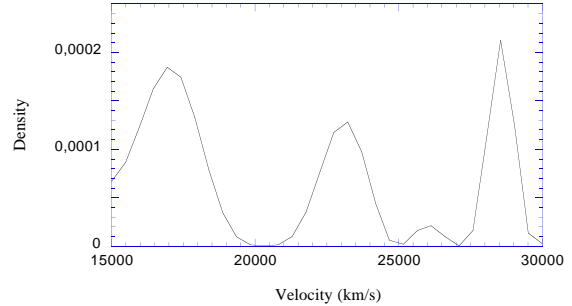}}
    \caption{A89 velocity distribution in the 15000-30000 \kms\ range.
        There are no galaxies with velocities smaller than 15000 \kms\ 
        in this sample.}
  \protect\label{vitS1}
\end{figure}

\begin{figure}[tbp] 
\centerline{ 
\psfig{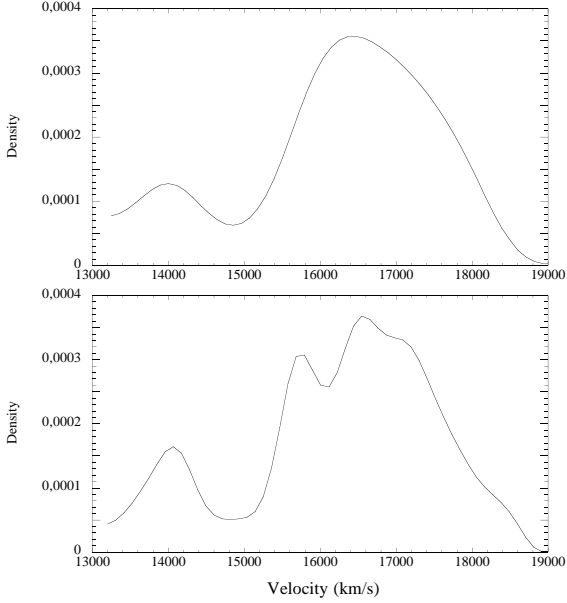}} 
\caption{Sc velocity distribution in the 13000-19000 \kms\ range. The two 
panels correspond to two velocity scales: top 110 \kms, bottom: 55 \kms.} 
\protect\label{vitSc} 
\end{figure}

\begin{figure}[tbp]
        \centerline{
        \psfig{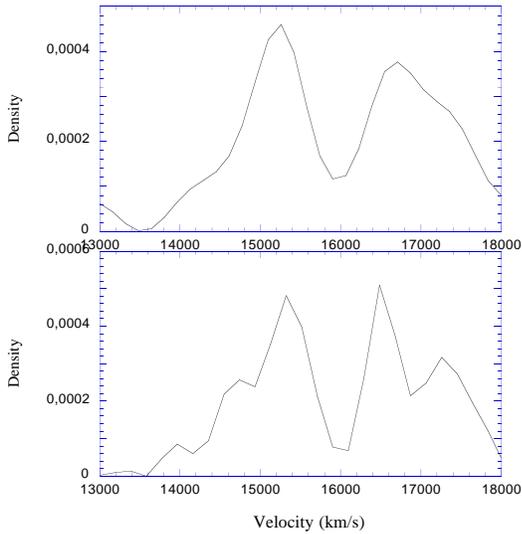}}
    \caption{A87 velocity distribution in the 13000-18000 \kms\ range. 
        The two panels correspond to two velocity scales: top 156 \kms,
bottom: 88 \kms.}
  \protect\label{vitS2}
\end{figure}

\begin{figure}[tbp]
        \centerline{
        \psfig{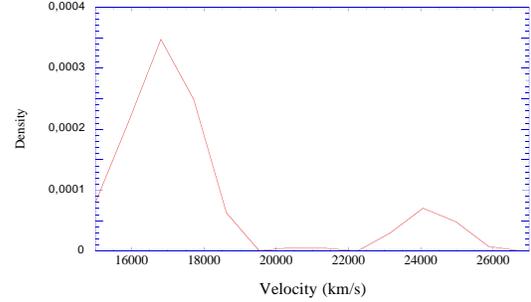}}
    \caption{C87 velocity distribution in the 15000-27000 \kms\ range.
        There is no velocity under 15000 \kms.}
\protect\label{vitS4}
\end{figure}

\begin{figure}[tbp]
        \centerline{
        \psfig{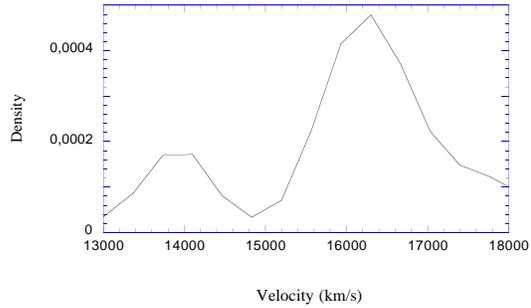}}
    \caption{C89 velocity distribution in the 13000-18000 \kms\ range.}
  \protect\label{vitS3}
\end{figure}

The velocity distributions in this range are displayed in
Figs.~\ref{vitS1}, \ref{vitSc}, \ref{vitS2}, \ref{vitS4} and
\ref{vitS3} for the A89, Sc, A87, C87 and C89 samples respectively.

We have analysed the distribution at various scales. However, 
due to the inhomogeneous number of galaxy velocities in the various 
samples distributed in various velocity intervals, the scales differ,
depending on the number of galaxies in each sample.

\subsubsection{A structure at ${\simeq 14000}$ \kms: the ``Foreground Group''}

A peak in the velocity distribution at 14000 \kms\ is observed for Sc
(Fig. \ref{vitSc}, top and bottom panels), A87 (Fig. \ref{vitS2},
bottom panel) and the comparison sample C89 (Fig.\ref{vitS3}).  This
peak corresponds to the ``Foreground Group'' detected by using the
Serna \& Gerbal (1996) hierarchical method and identified on the X-ray
image with the excess to the west of ABCG 85 (Durret et al. 1996).

\subsubsection{Structures in the [15000-19000 \kms] range}

In velocity space for the Sc and C89 samples, a dip at ${\simeq14800}$
\kms\ follows the \fg\ (see Figs.~\ref{vitSc} and \ref{vitS3}). Notice
that the galaxy samples A89, C87 and C1 contain no galaxy velocities
smaller than ${\simeq15000}$ \kms, while this is not the case for
the A87 sample (Fig.~\ref{vitS2}); in all these samples, most of the
galaxy velocities lie between 15000 and 19000 \kms.

In the case of Sc (top panel in Fig.~\ref{vitSc}), A89, C89, C87 and
for the two control samples C1 and C2, a maximum is observed at
$\simeq16500$ \kms. These velocity distributions are clearly not
Gaussian for any of the samples.
 
At a smaller scale, in the case of Sc, the velocity density breaks
into a two-maxima density distribution (Fig.~\ref{vitSc}, bottom
panel) with a first maximum at ${\simeq15700}$ \kms\ followed by a dip
(the \dip\ in the following) at ${\simeq16100}$ \kms, then a second maximum
at ${\simeq16800}$ \kms\ followed by a shoulder at
${\simeq17300}$~\kms. It is interesting to notice that the velocity of
this second maximum is very close to that of the cD galaxy.
 
Even if the relative levels of peaks and dips are not the same for
both samples, the succession of features in the Sc velocity density
distribution is very similar to that in the density profile of A87
(Fig.~\ref{vitS2}, top panel): a first maximum followed by the \dip\
then a second maximum.
        
In several samples we observe a similar succession of maxima and
minima, such as the 14000 \kms\ maximum, a small dip at 15000 \kms,
the \dip\ at 16100 \kms, and a maximum at
$\simeq16500$ \kms\ (Figs.~\ref{vitSc}, \ref{vitS2} and
\ref{vitS3}). A peak (at ${\simeq17300}$ \kms) corresponds in
Fig.~\ref{vitS2} (bottom panel) to the shoulder noted in the Sc sample
(Fig.~\ref{vitSc}, bottom panel).

\section{Three Dimensional Structure of the ABCG 85 Complex}\label{modeling}

\begin{figure}[tbp]
        \centerline{
        \psfig{figure=h0823.f13,height=7cm,clip=}}
    \caption{Galaxies in the right ascension - declination plane.
The colour coding is as follows: dark blue:~(13000\kms - 14000 \kms), 
green:~(14000\kms - 16000 \kms), purple:(16000\kms - 16800 \kms), 
red:~(16800\kms - 20000 \kms), yellow-green:~(20000\kms - 26000 \kms), 
cyan:~(26000\kms - 32000 \kms). }
\protect\label{xy}

\end{figure}

\begin{figure}[tbp]
        \centerline{
         \psfig{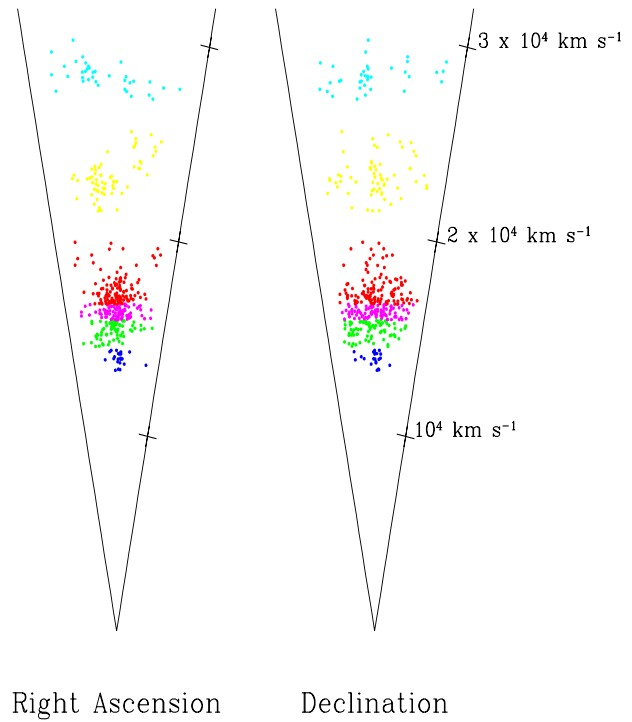}}
    \caption{ Cone diagram for the right ascension (left panel) and 
declination (right panel) of the ABCG 85/87/89 complex. The covered area
is 6000$\times$6000 arcsec$^2$. Galaxies with velocities lower than 13000
\kms\ have been removed from these plots.}
\protect\label{cone}

\end{figure}

\begin{figure}[tbp]
        \centerline{
        \psfig{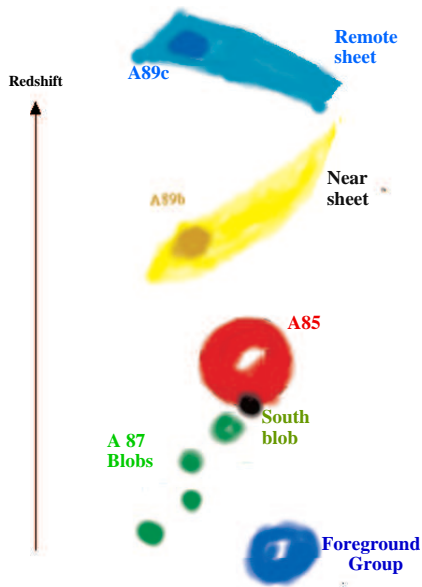}}
    \caption{An artist's view of the ABCG 85/87/89 complex. The dimensions 
are not to scale.}
\protect\label{artist}
\end{figure}

Figure \ref{xy} shows the distribution in the plane of the sky of
galaxies with measured velocities, colour coded for various velocity
ranges. This figure illustrates the difficulty in untangling structures
projected on the plane of the sky, even if they have very different
velocities. We can note however that two structures are visible: a
condensation of ``yellow galaxies'' in the region of ABCG 89 (but
galaxies in this velocity range also can be found elsewhere), and
``dark blue galaxies'' corresponding to the \fg. Besides, there seems
to be an enhancement of galaxies aligned along PA$\sim$160\deg.

We show two cone diagrams for the velocity distribution, which are
displayed in Fig.~\ref{cone}. Both panels show evidence for various
substructures along the line of sight.  The bulk of the galaxies (the
purple and red points in the figures) correspond to ABCG 85. 

In the left panel of Fig.~\ref{cone}, the \fg\ is clearly visible
(dark blue points), and it is apparent from Fig.~\ref{xy} that this
group is quite compact and concentrated (in projection) towards the
center-west of the cluster.

The ABCG 87 cluster itself, or rather the region south-east of ABCG
85, appears to be composed of several subgroups with velocities
somewhat lower than the median cluster velocity (green points). These
subgroups, defined in velocity space, have X-ray counterparts
(Fig.~\ref{excessx}). In Fig.~\ref{artist} we place each subgroup at
its X-ray position and represent it in the right ascension-velocity
plane (the velocities are in fact the various peaks observed in
Fig.~\ref{vitS2}), we observe a filament pointing towards ABCG 85.
This suggests that a stream of material is falling onto the main core
of ABCG 85. The south blob (the extended X-ray feature $10'$ south of
the cD) would then be the meeting point (see Fig.~\ref{excessx}) and
corresponds to a hot region in the ASCA temperature map (Markevitch et
al. 1998).

The two structures with the largest velocities (yellow and cyan
points) seem to trace two sheets of galaxies (referred to as the near
and remote sheets in Fig.~\ref{artist}). Both structures are inclined
along the line of sight, and appear to intersect.  These two sheets
may lie on the surface of a large bubble which intersects our field of
view. 

Note also the coalignment, along the line of sight, of the Foreground
Group, ABCG 85, A89b, and A89c. This could indicate the presence of a
second filament directly along the line of sight. Furthermore, the two
filaments could be concentrations in a sheet.  Notice the existence of
a dozen galaxies with velocities slightly higher than that for most of
the ABCG 85 galaxies which indicate a lower density region of the line
of sight filament.

\section{Discussion}

We presented a three dimensional model of the complex ABCG 85/87/89
region.  We find a general extension observed toward the south-east
along PA$\simeq 160$\deg\ both in X-rays and in optical photometric
observations:

- In X-rays, the elongated structure extends to the edge of the ROSAT
PSPC field of view; the X-ray appearance includes several individual
group-like systems lying up to about 4~Mpc (in projection) from the
center of the cluster (see Fig.~\ref{xcont}) as well as very elongated
diffuse emission;

- from the optical photometric catalogue, ABCG 85 is elongated
towards the ABCG 87 cluster (see Fig.~\ref{kernel}). The
PA$\simeq 160$\deg\ is characteristic for ABCG 85: the direction of
the major axis, the brightest galaxies in ABCG 85, and even the major
axis of the central cD galaxy itself are elongated along the same
direction.

East of ABCG 85 lies ABCG 89 which is not bright in X-rays. Our
velocity data reveal that ABCG 89 is not a cluster, but the
superposition on the sky of two groups well separated in velocity
space. These two groups (to which we refer as A89b \& A89c in
Fig.~\ref{artist}) are located in intersecting sheets on opposite
sides of a large bubble.

We have shown that ABCG 87 is not a rich cluster, but is resolved
into individual groups possibly falling onto ABCG 85.  These groups
are organized as a filament almost perpendicular to the plane of the
sky. The superposition of these groups gives the appearance of a
single optical cluster, while in X-rays, several groups are still
visible: This is probably due to the fact that the emissivity in
X-rays is proportional to the density squared, therefore enhancing the
contrast, compared to optical imaging. 

ABCG 85 itself is probably not fully relaxed, even if it appears
smooth and symmetric; the distribution of velocities is obviously not
gaussian and probably multi-modal.

In his scenario for the Coma cluster, West (1997) links the
orientations of Coma to the filament in which Coma is embedded. He
suggests that matter, including galaxies, groups and gas, falls onto
the cluster along this filament. The case of ABCG 85 reinforces this
cosmological scenario because \textit{we actually observe} the infall of
material (a filament of groups of galaxies and gas) onto ABCG
85. This result is consistent with the X-ray temperature map derived
from ASCA data by Markevitch et al.  (1998), which shows a temperature
enhancement in a region south of ABCG 85 and roughly perpendicular to
the general direction of 160\deg\ along which the various structures
are aligned. Such a temperature increase could be interpreted as shock
heating due
to the compression of the X-ray gas by infalling matter.
The fact that there also appears to be a radio relic, also
roughly perpendicular to the 160\deg\ direction, in this zone (Bagchi
et al. 1998) also supports recent merger activity,
since  relativisitic electrons can be produced during a merger.

Remarkably, the ABCG 85/87 filament is coaligned with a much larger
structure including from northwest to southeast: ABCG 70, ABCG 85 and
89, ABCG 87, ABCG 91, the NGC 255 group and ABCG 106 (Fig.~1 in Slezak
et al. 1998). Such a structure extends over more than 5\deg\ on the
sky, corresponding to a linear distance of 28 Mpc at the redshift of
ABCG 85 (z=0.0555). This is obviously a lower
limit, since the filament may be inclined to the line of sight.
Although this projected value is smaller than that found for example
for the Perseus Pisces structure (50$h^{-1}$ Mpc, Haynes \& Giovanelli
1986), it is nevertheless much larger than typical cluster sizes.
Unfortunately, there are only very few redshifts available for this large
structure outside the ABCG 85/87/89 complex. Further optical
observations are required to confirm the cohesiveness of this
remarkable filamentary structure.

\begin{acknowledgements}
We are very grateful to A.~Biviano, D.~Fadda and E.~Slezak for making 
their software available to us. W. Forman and C. Jones thank the
IAP for its hospitality and acknowledge support from the Smithsonian
Institution and NASA contract NAS8-39073.
\end{acknowledgements}

\end{document}